\documentclass[conference]{IEEEtran}
\IEEEoverridecommandlockouts
\usepackage{cite}
\usepackage{amsmath,amssymb,amsfonts}
\usepackage{algorithmic}
\usepackage{graphicx}
\usepackage{titlesec}
\usepackage{enumerate}
\usepackage[sort&compress,numbers]{natbib}
\usepackage{svg}
\usepackage{xcolor}
\usepackage{float}
\usepackage{subcaption}
\usepackage{textcomp}
\usepackage{chngcntr}
\usepackage{graphicx}
\counterwithout{figure}{section}
\graphicspath{{images/}} 
\usepackage{xcolor}
\counterwithout{equation}{section}

\newcommand\extralabel[2]{\edef\@currentlabel{\@currentlabel#2}\label{#1}}
\makeatletter
\newcommand{\linebreakand}{%
  \end{@IEEEauthorhalign}
  \hfill\mbox{}\par
  \mbox{}\hfill\begin{@IEEEauthorhalign}
}
\def\BibTeX{{\rm B\kern-.05em{\sc i\kern-.025em b}\kern-.08em
    T\kern-.1667em\lower.7ex\hbox{E}\kern-.125emX}}
\makeatother

\begin{document}
    \title{An Optimized H5 Hysteresis Current Control with Clamped Diodes in Transformer-less Grid-PV Inverter \\}

\author{\IEEEauthorblockN{1\textsuperscript{st} Sushil Phuyal}
\IEEEauthorblockA{\textit{Department of Electrical Engineering} \\
\textit{Tribhuvan University}\\
Lalitpur, Nepal \\
076bel047.sushil@pcampus.edu.np}

\and
\IEEEauthorblockN{2\textsuperscript{nd} Shashwot Shrestha}
\IEEEauthorblockA{\textit{Department of Electrical Engineering} \\
\textit{Tribhuvan University}\\
Lalitpur, Nepal \\
076bel043.shashwot@pcampus.edu.np}

\and
\IEEEauthorblockN{3\textsuperscript{rd} Swodesh Sharma}
\IEEEauthorblockA{\textit{Department of Electrical Engineering} \\
\textit{Tribhuvan University}\\
Lalitpur, Nepal \\
076bel048.swodesh@pcampus.edu.np}
\and
\IEEEauthorblockN{4\textsuperscript{th} Rachana Subedi}
\IEEEauthorblockA{\textit{Department of Electrical Engineering} \\
\textit{Tribhuvan University}\\
Lalitpur, Nepal \\
076bel033.rachana@pcampus.edu.np}

\and
\IEEEauthorblockN{5\textsuperscript{th} Anil Kumar Panjiyar}
\IEEEauthorblockA{\textit{Department of Electrical Engineering} \\
\textit{Tribhuvan University}\\
Lalitpur, Nepal \\
anil.panjiyar@pcampus.edu.np}

\and
\IEEEauthorblockN{6\textsuperscript{th} Mukesh Gautam}
\IEEEauthorblockA{\textit{Electricity Infrastructure \& Buildings Division} \\
\textit{Pacific Northwest National Laboratory}\\
 Richland, Washington, USA \\
mukesh.gautam@pnnl.gov}

}

\maketitle

    \begin{abstract}
With the rise of renewable energy penetration in the grid, photovoltaic (PV) panels are connected to the grid via inverters to supply solar energy. Transformer-less grid-tied PV inverters are gaining popularity because of their improved efficiency, reduced size, and lower costs. However, they can induce a path for leakage currents between the PV and the grid part due to the absence of galvanic isolation between them. This leads to serious electromagnetic interference, loss in efficiency and safety concerns. The leakage current is primarily influenced by the nature of the common mode voltage (CMV), which is determined by the switching techniques of the inverter. 
In this paper, a novel inverter topology of  Hysteresis Controlled H5 with Two Clamping Diodes(HCH5-D2) has been derived. The HCH5-D2 topology helps to decouple the AC part (Grid) and DC part (PV) during the freewheeling to make the CMV constant and in turn, reduces the leakage current. Also, the additional diodes help to reduce the voltage spikes generated during the freewheeling period and maintain the CMV at a constant value. Finally, a 2.2kW grid-connected single-phase HCH5-D2 PV inverter system's MATLAB simulation has been presented with better results when compared with a traditional H4 inverter.
\end{abstract}


    \section{Introduction}
Distributive Energy Resources (DERs) consisting of a variety of energy types such as solar, wind, and Battery Energy Storage System (BESS) are generally connected to a centralized or islanded power grid. DERs have been widely adopted in both commercial and residential areas \cite{awasthi2017energy}. Also, DERs such as solar energy and wind energy are extensively used to complement BESS, enabling the storage of excess energy during low demand and its utilization during high demand \cite{phuyal2023predictive}. 
\\
During periods of sufficient sunlight, grid integrated inverters are used to inject or store solar energy by converting DC power into AC and involve a feedback loop \cite{vakacharla2020state}. Line frequency transformers are mostly used in commercial PV inverters to provide galvanic isolation between PV and grid but are often large, heavy, and expensive. Transformer-less inverters (TLI) are being developed to increase efficiency, reduce size, and lower costs \cite{shayestegan2018overview}. TLIs exhibit higher efficiency due to the absence of losses associated with magnetic coupling, including core and copper losses \cite{9121749}. However, removing the isolation capability of the transformer requires careful consideration to ensure safety and reliability when connecting solar power directly to the grid \cite{biswas2022new}. Grounding the PV frame to earth introduces parasitic capacitances between the PV array and the ground \cite{pourmirasghariyan2022dc}, which range from 60 nF/kW to 160 nF/kW in normal conditions. This parasitic capacitance forms an LC resonant circuit consisting of a PV array, grid, and converter circuit, allowing unwanted leakage current to flow into the grid, leading to harmonic content, losses, and electromagnetic interference. The German standard VDE0126-1-1 states that leakage currents over 300mA must trigger a break within 0.3 seconds and send a fault signal \cite{lcstandardcite}.
\\
The leakage current primarily depends upon the magnitude and nature of CMV. This CMV is the function of switching pattern of the inverter switches. In conventional inverter topologies, the CMV is not constant, as a consequence, this leads to the injection of leakage current to the inverter and to the grid too \cite{lccite}. There are lots of inverter topologies that use various combinations of switches that makes the CMV constant or DC in nature. Bipolar modulation can maintain constant CMV but due to its bipolar nature the switching losses of the inverter is high where as unipolar modulation is able to reduce the switching losses but cannot maintain constant CMV \cite{blaacha2021comparative}. So, using H4 topology for leakage reduction with proper efficiency is not possible \cite{H4conventional}. Hence, the main aim for controlling the inverter is to suitably mix both modulation techniques as proposed in HCH5-D2.  
\\
Various inverter topologies aim to maintain a constant CMV and improve differential-mode voltage (DMV) performance. Decoupling the AC side from the DC side during freewheeling periods is one effective method for weakening CMV. Topologies such as H5 \cite{h5cite} (DC side decoupling), H6 \cite{h6cite}, and HERIC \cite{hericcite} (AC side decoupling) have been developed to reduce CMV fluctuations. However, simple decoupling leads to CMV oscillations at resonant frequencies, making it unstable during freewheeling. Clamping circuits have been introduced to stabilize the CMV during these periods \cite{dH5cite}. While previous studies laid the groundwork for transformer-less PV inverters, newer topologies like HCH5-D2 require experimental validation under diverse grid conditions. This paper focuses on common-mode resonant circuit modeling of the HCH5-D2 inverter using the back-to-back source transformation method (Section II), discusses the inverter's modes of operation (Section III), introduces dual control techniques (Section IV), presents simulation results using Matlab/Simulink (Section V), and concludes with key findings (Section VI).

    \section{Proposed HCH5-D2 Inverter Topology}
\label{lc_analysis}
\begin{figure}[h]
    \centering
    \includegraphics[width = 0.7\linewidth]{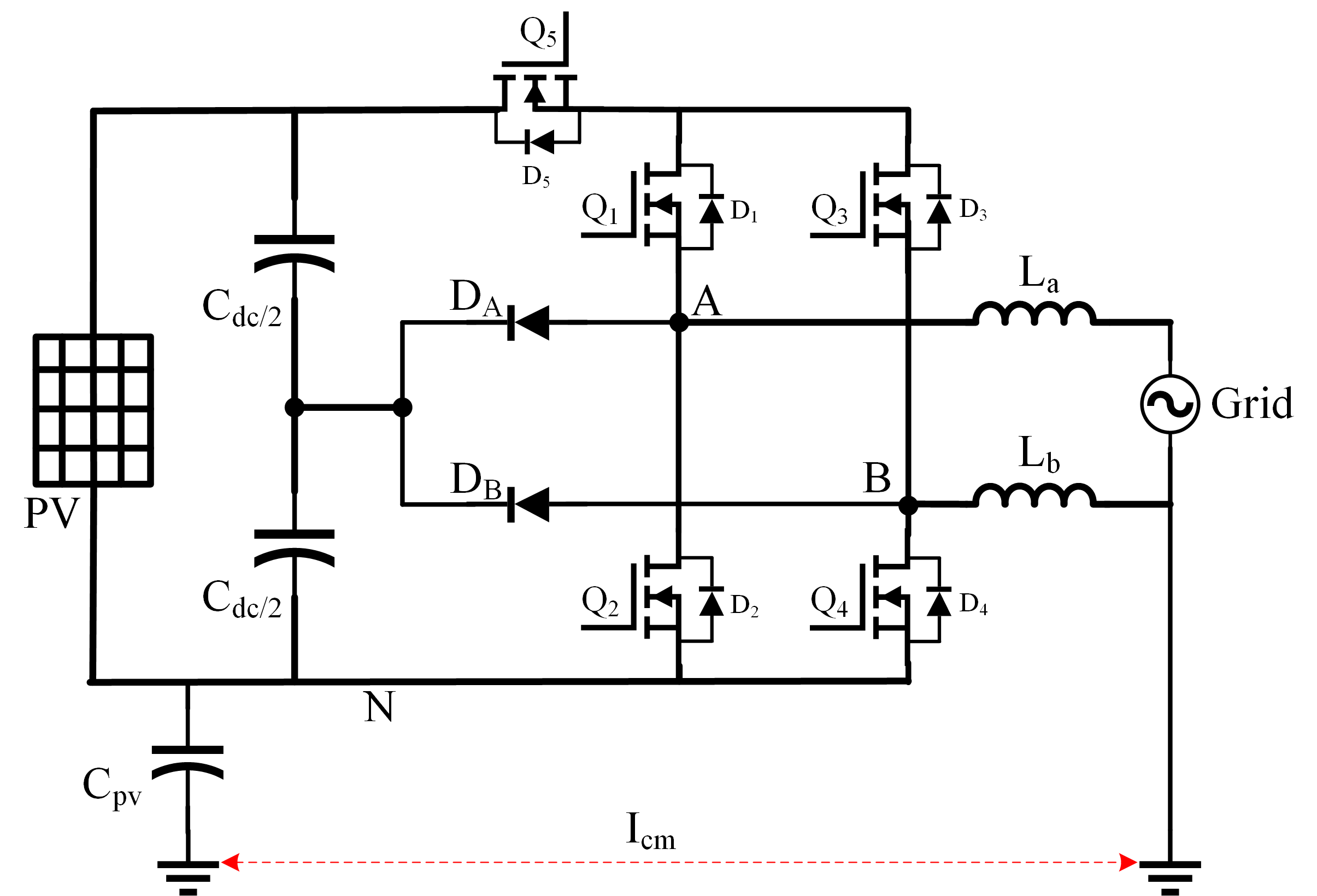}
    \caption{Main Circuit of HCH5-D2 Inverter Topology}
    \label{fig:h5main}
\end{figure}

The proposed HCH5-D2 inverter topology uses an additional power switch Q5 compared to H-bridge circuit as displayed in Fig \ref{fig:h5main}. Q5 is placed on the photovoltaic (PV) side, giving this converter a DC-decoupling type of TLI configuration. Q5 is responsible for isolating and disconnecting the DC or PV side of the circuit from the main grid by turning off when necessary thus achieving DC decoupling functionality. Additionally, two clamping diodes are added as shown in Fig.\ref{fig:h5main}, whose function is to clamp a CMV to constant magnitude during the freewheeling period. Therefore, these diodes help to reduce the CMV spikes created during the freewheeling period and maintain it to a constant value \cite{dhara2021nine}.
Stray capacitive effects between the photovoltaic (PV) array terminals and electrical ground are represented by the parasitic capacitance $C_{pv}$ in the Fig \ref{fig:h5main}. These parasitic capacitances demonstrate the distributed impact of capacitive coupling to stray fields. The pole voltages $V_{AN}$ and $V_{BN}$ at the positive and negative terminals of the PV array are determined by the switching states of the MOSFETs in the converter topology. Specifically, if the upper MOSFETs Q1, Q3, and Q5 are turned on, the pole voltage will be equal to the DC link voltage $V_{DC}$. Conversely, if the lower MOSFETs Q2 and Q4 are conducting, the pole voltages will be zero volts.

\begin{figure}[h]
    \centering
    \begin{subfigure}[b]{0.45\linewidth}
        \centering
        \includegraphics[width=\linewidth]{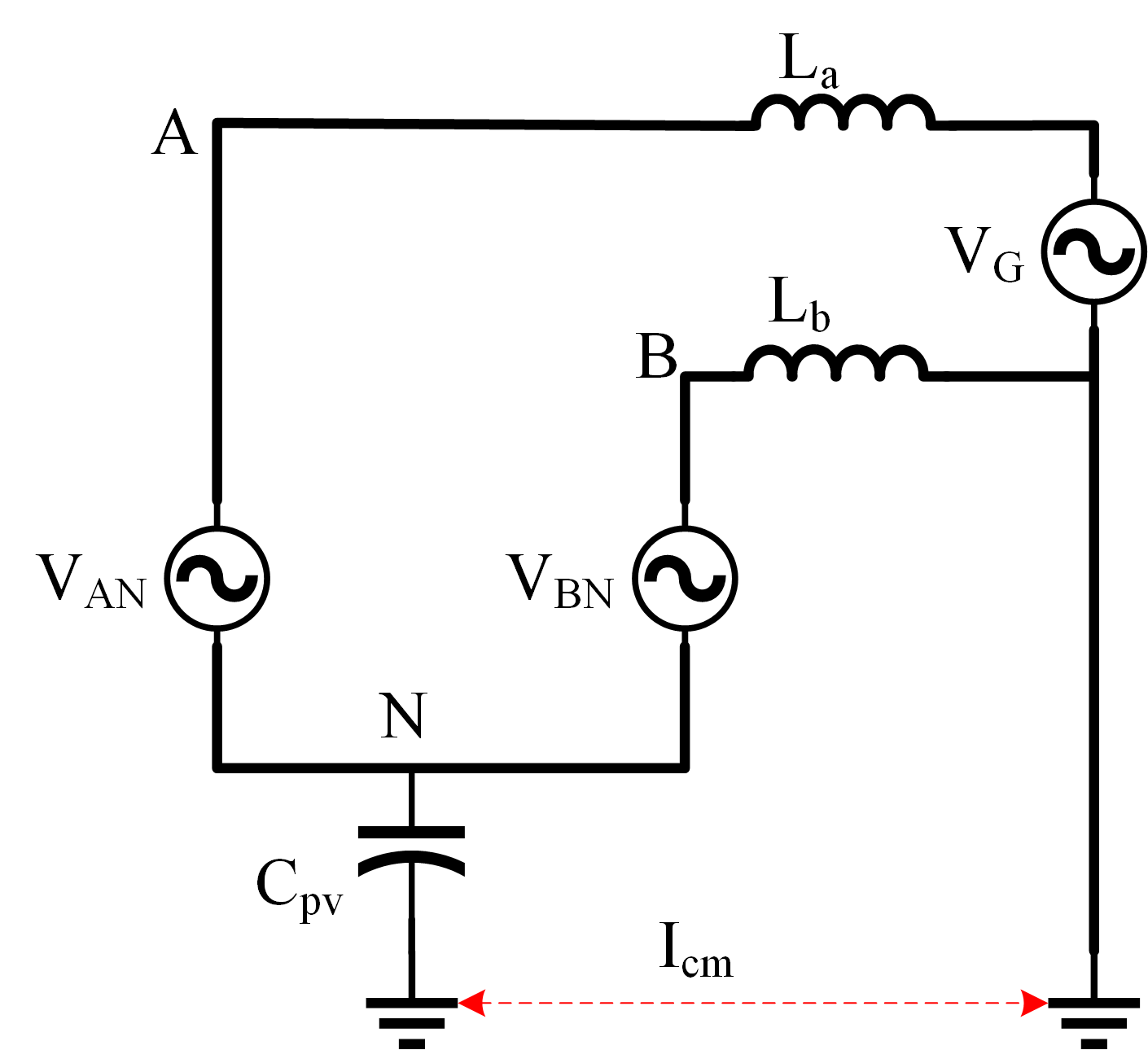}
        \caption{Resonant circuit in terms of pole voltages}
        \label{fig:h5pole}
    \end{subfigure}%
    \begin{subfigure}[b]{0.45\linewidth}
        \centering
        \includegraphics[width=\linewidth]{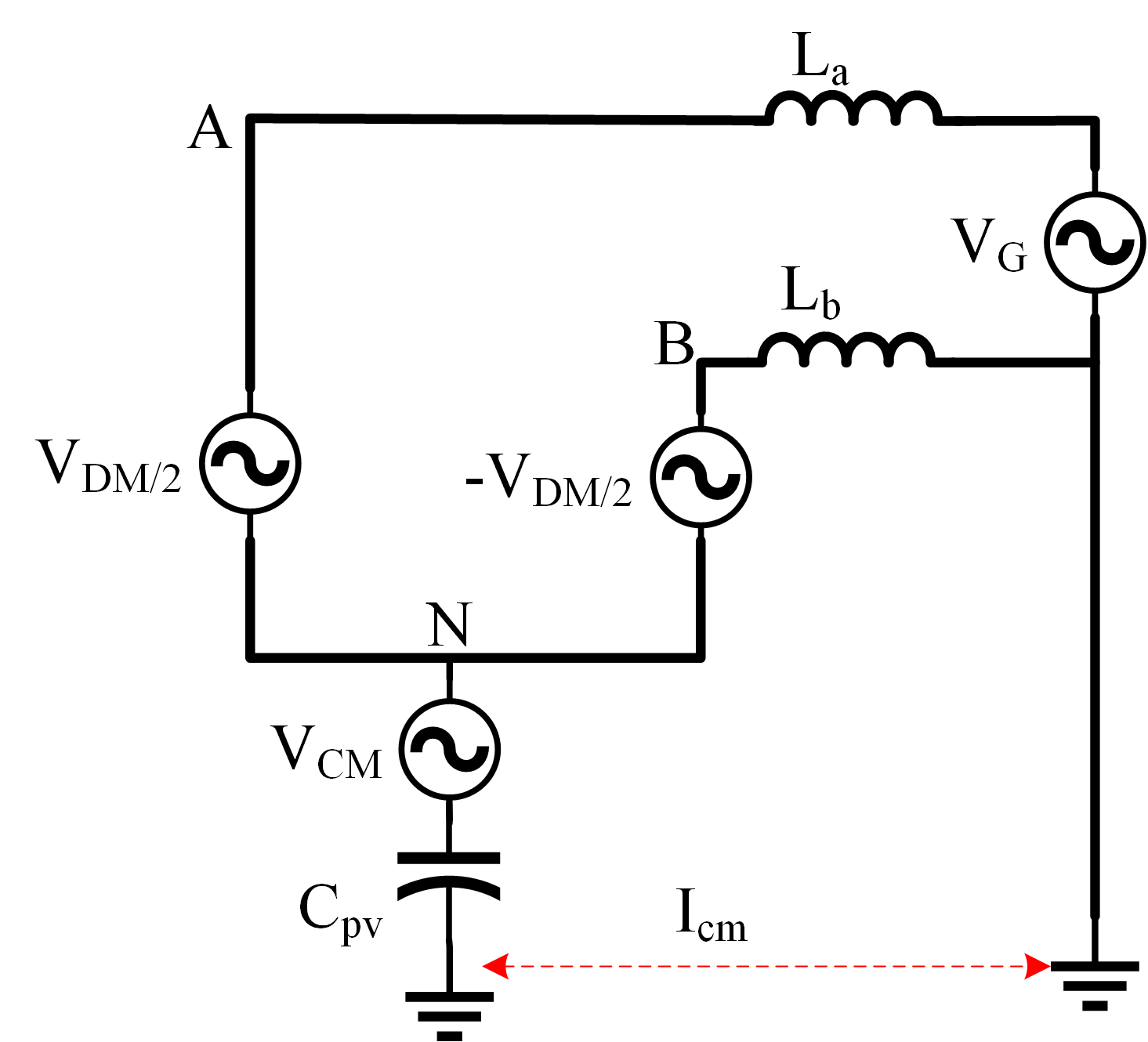}
        \caption{Resonant circuit in terms of DM and CM voltages}
        \label{fig:h5pole2cmdm}
    \end{subfigure}
    \caption{Resonant circuits of H5 inverter}
    \label{fig:h5_resonant_circuits}
\end{figure}

Likewise Fig.\ref{fig:h5main} illustrates the path of the leakage current $I_{cm}$, which flows through both the phase and neutral lines of the inverter and to the grid. The modulation strategy employed in this hysteresis current control inverter topology produces a combination of common mode (CM) and differential mode (DM) characteristics when generating the inverter switching signals. The improved CM characteristics' objective is to keep the CMV constant, which reduces the leakage current. Better DM characteristics allow the circuit to generate a multilevel DMV or inverter voltage. As shown in Fig.\ref{fig:h5pole}, the circuit in Fig.\ref{fig:h5main} can be resolved into its pole voltages $V_{AN}$ and $V_{BN}$.

\par
The average of the pole voltages is known as the CMV. Similarly, the difference between the pole voltages is known as DMV. CMV and DMV can be expressed as.
\begin{equation}\label{eqn:cmv}
   CMV (V_{CM})=\frac{V_{AN}+V_{BN}}{2} 
\end{equation}

\begin{equation}\label{eqn:dmv}
   DMV (V_{DM})=V_{AN}-V_{BN} 
\end{equation}
Using the equations \eqref{eqn:cmv} and \eqref{eqn:dmv}, the pole voltages $V_{AN}$ and $V_{BN}$  can be expressed in terms of $V_{CM}$ and $V_{DM}$ as 
\begin{equation}\label{eqn:pole2cmdm}
\begin{aligned}
& V_{A N}=\mathrm{V}_{\mathrm{CM}}+\dfrac{V_{D M}}{2} \\
& V_{B N}=\mathrm{V}_{\mathrm{CM}}-\dfrac{V_{D M}}{2} 
\end{aligned}
\end{equation}
Likewise, the equivalent circuit is shown in Fig.\ref{fig:h5pole} and it can be re-expressed in terms of CM and DM voltage according to \eqref{eqn:pole2cmdm} as in Fig.\ref{fig:h5pole2cmdm}. A comprehensive analysis is shown below to examine the behavior of leakage current in a transformer-less inverter\cite{h4maths}. 
\subsection{Conversion of Voltage Sources into Current Sources}
\begin{figure}[!h]
    \centering
    \includegraphics[width = 0.45\linewidth]{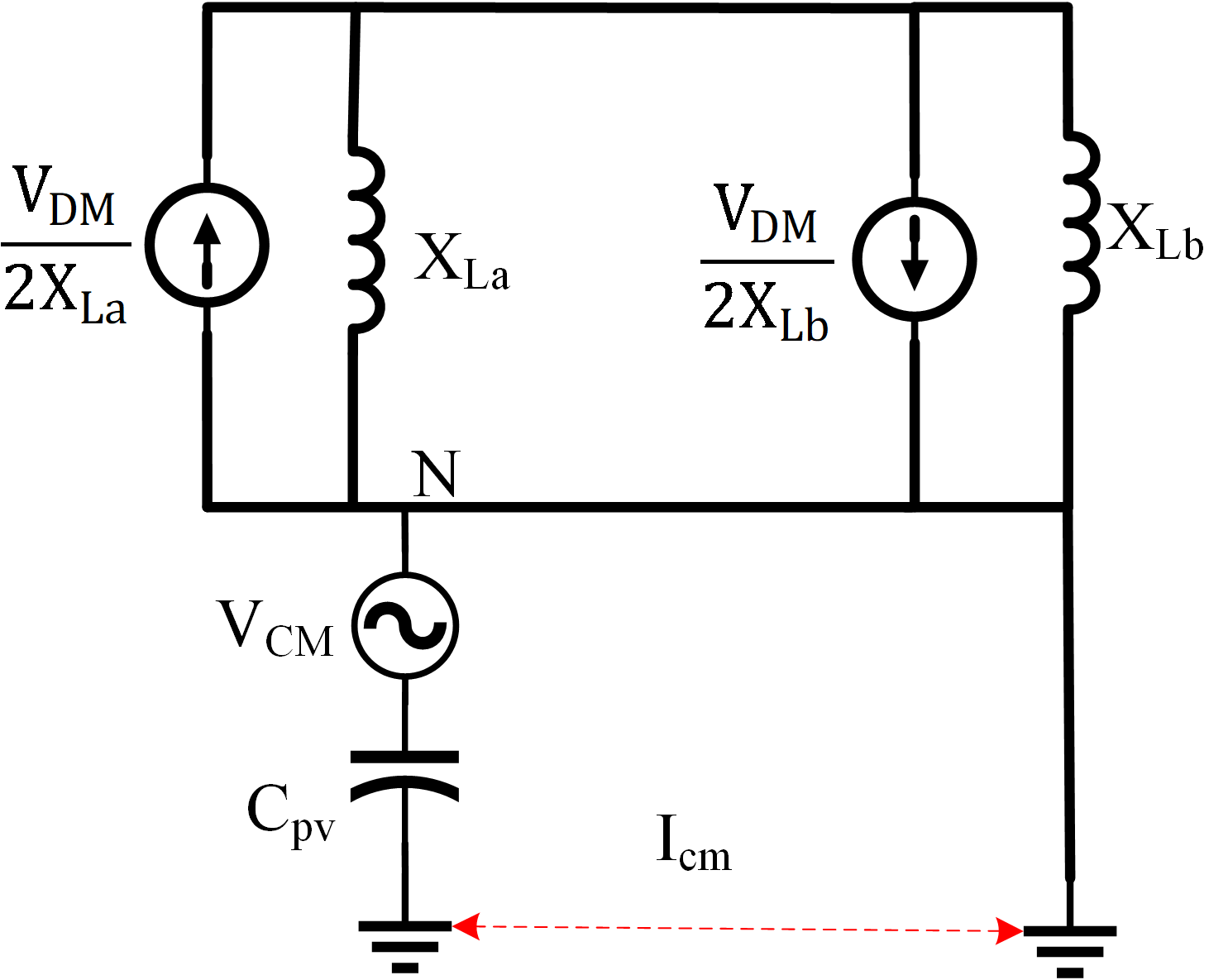}
    \caption{Resonant circuit of H5 inverter with current sources}
    \label{fig:h5_nogrid}
\end{figure}

The voltage sources which is in series with reactance can be transformed into their respective current sources as displayed in Fig\ref{fig:h5_nogrid}. Both current sources are operating at very high switching frequencies in the order of kHz whereas grid voltage frequency is just 50 Hz. Therefore, for high frequency analysis, the grid voltage can be neglected.

The two current sources of Fig.\ref{fig:h5_nogrid} in parallel connection can be further resolved into a single current source and with a single voltage source as shown in Fig.\ref{fig:h5_1curr}.


\subsection{Conversion of Current Source into Voltage Source}
\begin{figure}[h]
    \centering
    \begin{subfigure}[b]{0.45\linewidth}
        \centering
        \includegraphics[width=\linewidth]{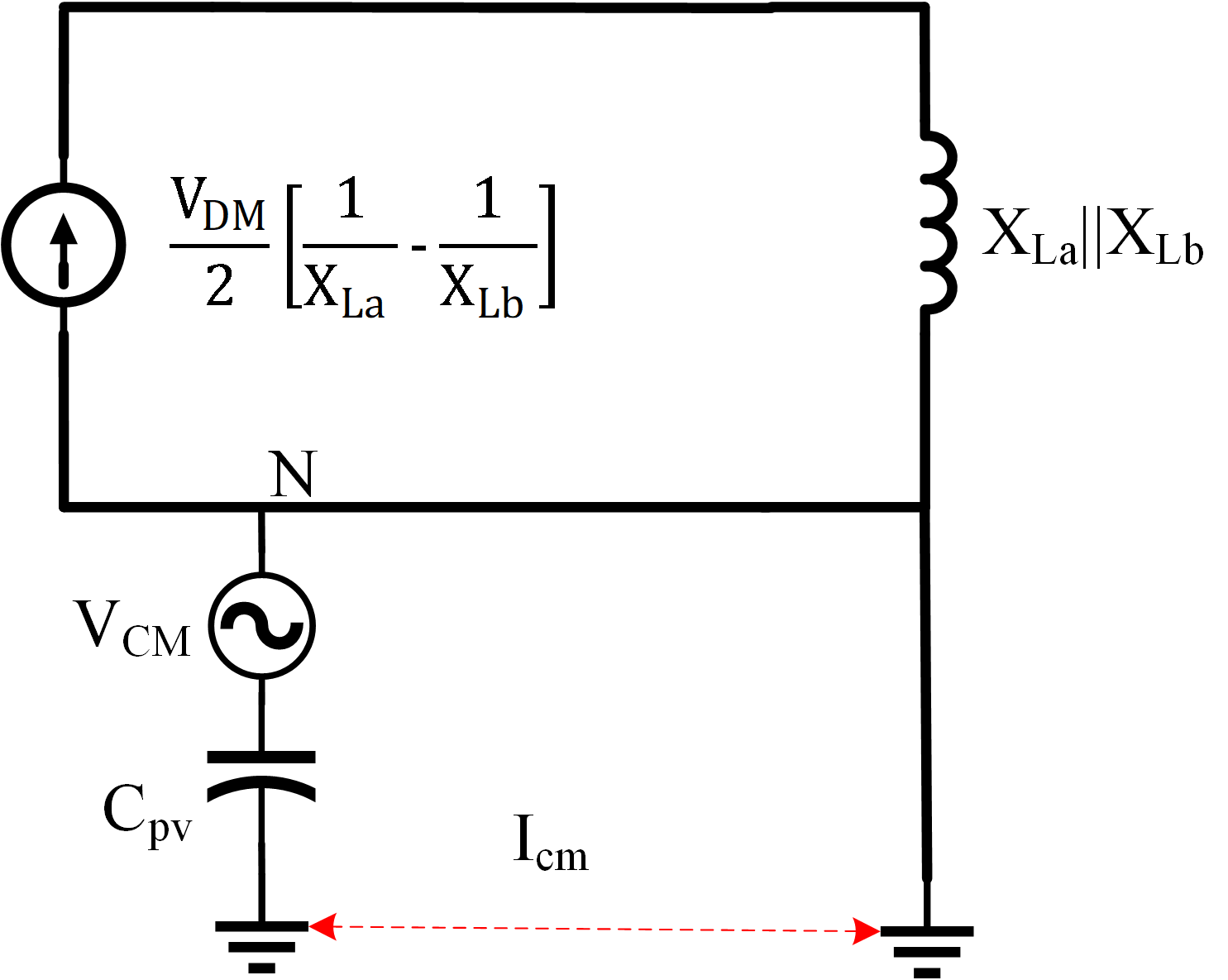}
        \caption{A single voltage and current source}
        \label{fig:h5_1curr}
    \end{subfigure}%
    \begin{subfigure}[b]{0.40\linewidth}
        \centering
        \includegraphics[width=\linewidth]{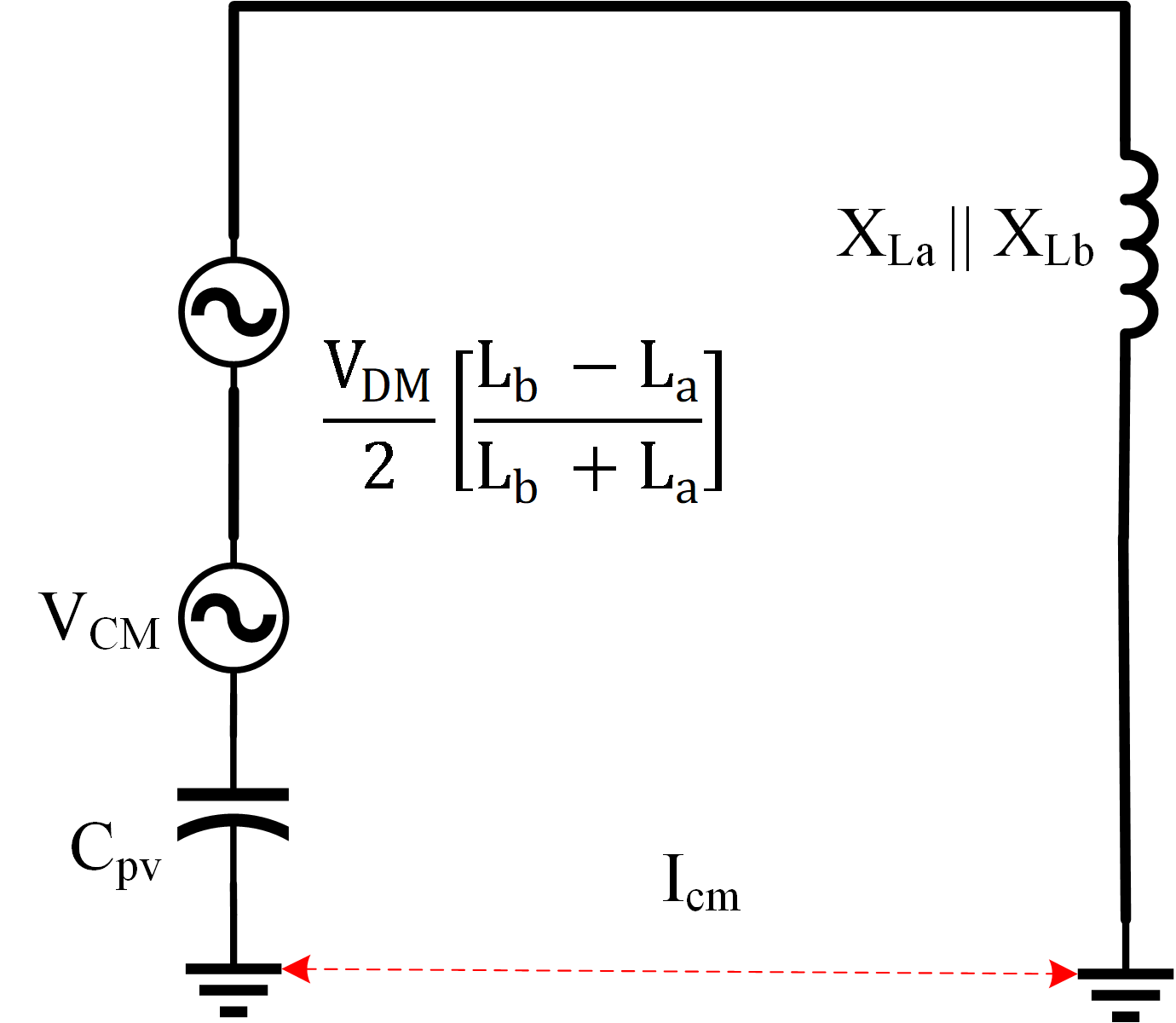}
        \caption{In the form of two voltage sources}
        \label{fig:h5_2volt}
    \end{subfigure}
    \caption{Resonant circuits of H5 inverter in various voltage and current source configurations}
    \label{fig:h5_resonant_voltage_current}
\end{figure}

The single current source in parallel with the reactance in Fig.\ref{fig:h5_1curr} is transferred back to the voltage source with a series reactance, as shown in Fig.\ref{fig:h5_2volt}.


The two voltage sources can be added to form total CMV, $V_{tCM}$\cite{estevez2020leakage}. The final equivalent circuit, to properly describe the nature of CM current is shown in Fig.\ref{fig:h5_1volt}
\begin{equation}\label{eqn:totalcmv}
    V_{tCM} = V_{CM} + \dfrac{V_{DM}}{2} \left( \dfrac{L_B - L_A}{L_B + L_A} \right)
\end{equation}
From \eqref{eqn:totalcmv}, it can be understood that when the value of filter inductors $L_A$ and $L_B$ are equal, in both phase and neutral of the grid side, then $V_{tCM}$ will be equal to $V_{CM}$ only with no effect of DM voltage.
\subsection{Leakage Current and Resonant Frequency}
\begin{figure}[!h]
    \centering
    \includegraphics[width = 0.4\linewidth]{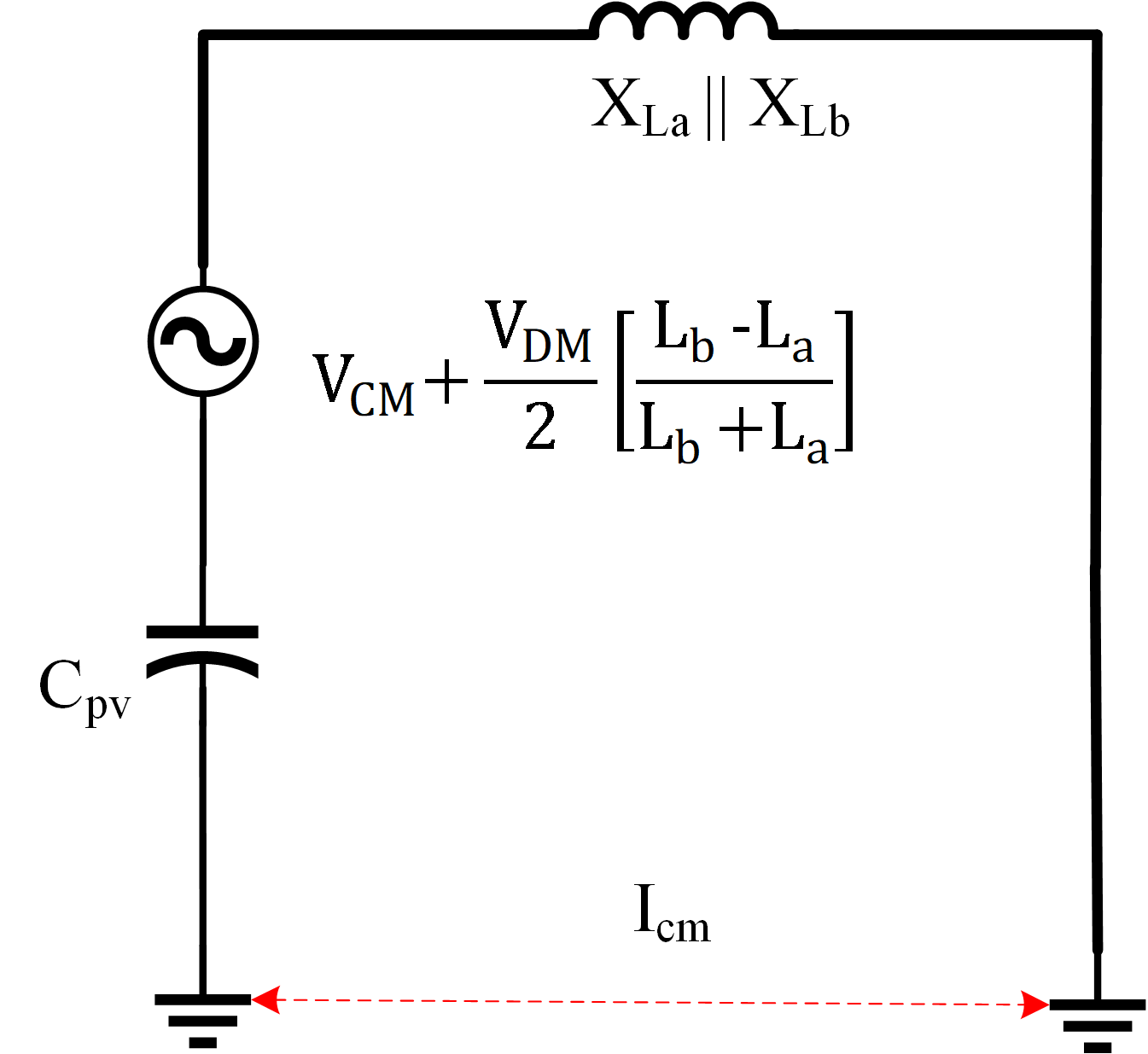}
    \caption{Simplified resonant circuit of H5 inverter in the form of single common-mode voltage source}
    \label{fig:h5_1volt}
\end{figure}

From Fig.\ref{fig:h5_1volt}, the expression of the leakage current is expressed as
\begin{equation}\label{eqn:lc}
    I_{CM} = \dfrac{V_{tCM}}{(X_{LA}||X_{LB})+X_{C_{PV}}}
\end{equation}
From \eqref{eqn:lc}, it is concluded that the flow and magnitude of leakage current depend upon the value and nature (resonating frequency) of CMV.
\par
The equivalent circuit is called an LC resonant circuit due to the formation of two energy storage elements i.e. equivalent leakage capacitance, $C_{eqPV}$ and equivalent filter inductance, $L_{eq}$ = $L_A$ $||$ $L_B$. The LC circuit will oscillate at a resonance frequency given by the equation \eqref{eqn:resfreq}.
\begin{equation}\label{eqn:resfreq}
    F_{resonant} = \dfrac{1}{2\pi} \sqrt{\dfrac{1}{L_{eqv}C_{eqv}}}
\end{equation}
Equation \eqref{eqn:lc}, proves that leakage current depends upon the magnitude as well as the nature of CMV. When the CMV is constant (DC) for each modes of operation, then it will be a DC quantity with a resonant frequency being zero. Therefore, the whole of the denominator of \eqref{eqn:lc} becomes infinity, and hence leakage current tends to zero.
\\

    \section{Modes of Operation in Proposed HCH5-D2 Inverter}
Unlike a standard H5 topology where the CMV at the freewheeling period is uncertain \cite{h5cite}, in the proposed HCH5-D2 topology, the two diodes clamps the pole voltages $V_{AN}$ and $V_{BN}$ to $\dfrac{V_{DC}}{2}$. The upper diode clamps the upper pole voltage and the lower diode clamps the lower pole voltage to $\dfrac{V_{DC}}{2}$ during each of the freewheeling modes. The power transfer in the proposed topology is completed in following four modes of operation.

\begin{figure}[h]
    \centering
    \includegraphics[width=1\linewidth]{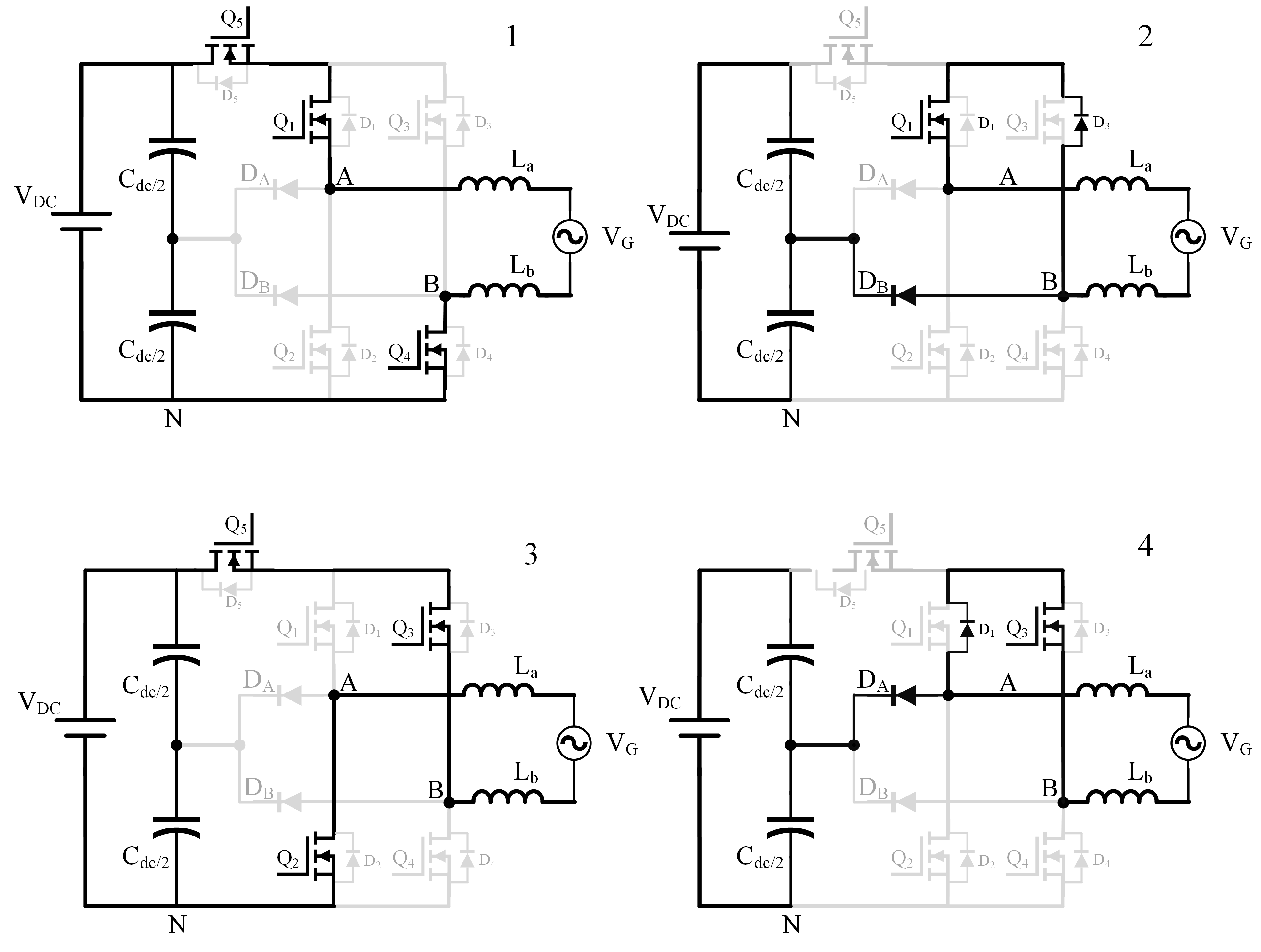}
    \caption{Different Modes of Operation of HCH5D2}
    \label{fig_modes}
\end{figure}

The switching network in Mode 1 allows energy to flow through switches Q1, Q4, and Q5, from the PV to the grid, to get an output of $V_{DC}$; the CMV of this mode is equal to $V_{CM}$ = $V_{DC}/2$. For Mode 2, all the switches, except Q1, are turned on, and current freewheels through Q1 and diode D3, isolating the DC side from the AC side. The clamping circuit sets the pole voltage $V_{BN}$ to $V_{DC}/2$ which results in zero output voltage, while $V_{CM}$ remains $V_{DC}/2$. In Mode 3, only switches Q2, Q3, and Q5 are on which transfer energy and generate an output voltage of $-V_{DC}$. Here, the CMV still at $V_{CM}$ = $V_{DC}/2$. Finally, in Mode 4, all switches except Q3 are closed, and the current freewheels through Q3 and diode D1, again decoupling the DC and AC sides. The clamping circuit sets $V_{AN}$ to $V_{DC}/2$, with zero output voltage and the CMV remaining at $V_{DC}/2$. The four modes are illustrated together in Fig. \ref{fig_modes}.

Based on these four modes, the pole voltages $V_{AN}$ and  $V_{BN}$  along with the CM and DM voltages are depicted in the Table \ref{tab:h5tab}. It is evident that DM voltage maintains unipolar performance i.e. three level voltages $+V_{pv}$, 0, and $-V_{pv}$. Likewise, CM voltage has a constant value in all of the modes. During freewheeling mode, a diode clamps the pole voltages to $V_{pv}/2$. As a result, CM voltage remains constant i.e. $V_{pv}/2$ in all of the modes of operation.
\begin{table}[!h]
    \centering
    \caption{CMV and DMV in different modes of operation of HCH5-D2 topology.}
    \label{tab:h5tab}
\begin{tabular}{p{1cm}|p{1cm}|p{1cm}|p{1cm}|p{1cm}}
\hline Modes & $V_{AN}$ & $V_{BN}$ & $V_{DM}$ & $V_{CM}$ \\
\hline
(a) & $V_{DC}$ & 0 & $V_{DC}$ & $V_{DC}/2$ \\
(b) & $V_{DC}/2$ & $V_{DC}/2$ & 0 & $V_{DC}/2$ \\
(c) & 0 & $V_{DC}$ & $-V_{DC}$ & $V_{DC} / 2$ \\
(d) & $V_{DC}/2$ & $V_{DC}/2$ & 0 & $V_{DC}/2$\\
\hline
\end{tabular}
\end{table}

\section{Dual Loop Control Strategy for the proposed HCH5-D2 Inveter}
\begin{figure}[h]
    \centering
    \includegraphics[width=\linewidth]{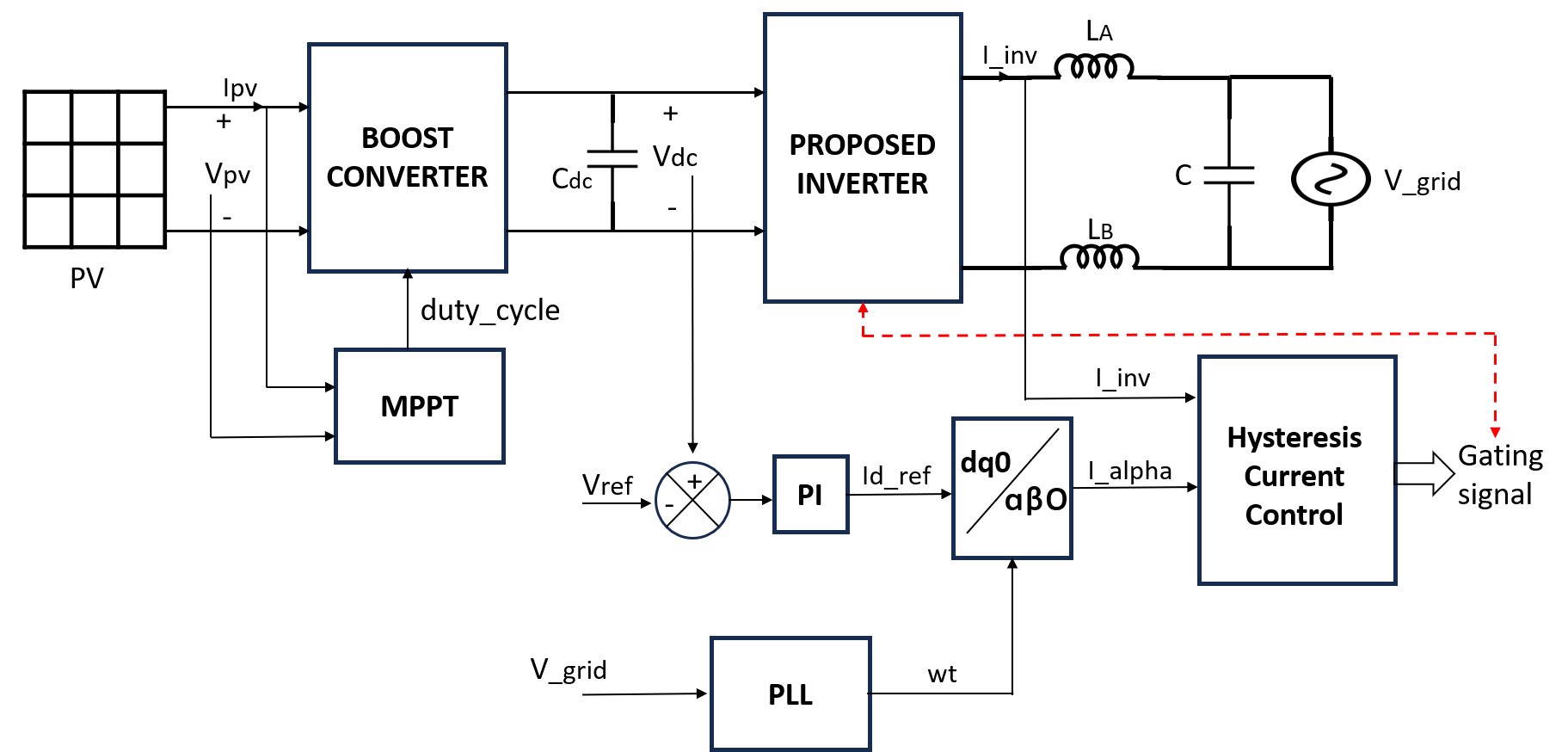}
    \caption{A control structure diagram for HCH5D2 inverter system.}
    \label{fig_h5control}
\end{figure}

The Fig \ref{fig_h5control} shows the structure for the control strategy that is implemented for HCH5-D2. The system consists of Photovoltaic Panel, Boost converter with MPPT system, HCH5-D2 inverter, and the EMI filters.Voltage control loop and current control loop are included in the control structure, which are explained in the following subsequent sections.

\subsection{Outer DC Link Voltage Control Loop}
To extract maximum power from PV panel, Perturb and Observation (P\&O) algorithm is used. The step up boost converter regulates and adjusts the output voltage of the PV panel to match the target voltage, resulting in maximum power production.  This loop regulates DC-link voltage at a constant level. Here, the reference voltage value is set as 400 V. By adjusting the reference DC current Id\_ref, voltage regulation is achieved. It should be noted that the current Id\_ref represents the active component of the reference grid current which corresponds to the active power available at the DC side. 

\subsection{Hysteresis Current Control (Inner Grid)}
\begin{figure}[H]
    \centering
    \includegraphics[width = 0.65\linewidth]{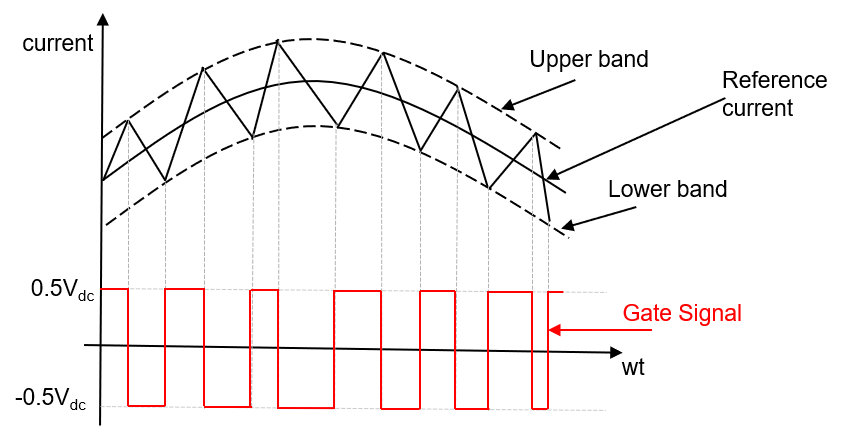}
    \caption{A Visual Representation of Hysteresis Band Current Control.}
    \label{fig_hcc_graph}
\end{figure}

This loop's main function is to synchronize the inverter with the grid i.e., to inject active power to the grid and help to maintain the voltage of DC-link at a fixed value.  
With the Inverse Park transformation, the alpha component of the current corresponding to Id\_ref is used as a reference current in order to set the inverter current to a desirable value. This is achieved from  the hysteresis band current controller (HBCC) loop\cite{bode2000implementation}. HBCC compares I\_inv and I\_alpha in order to set I\_inv to a desirable value.  The magnitude of the actual current (I\_inv) is controlled by controlling gating signals of the inverter. Fig.\ref{fig_hcc_graph} shows the variation gate signal in order to maintain the actual current within the tolerance level.    
    \section{Results}
Proposed HCH5D2 inverter's simulation is performed in MATLAB(Simulink) software with parameters as shown in Table \ref{tab_sim}. The proposed HCH5D2 topology of inverter was analysed and compared with the conventional H4 topology. Grid voltage has been obtained to be in phase with grid current, and hence unity power factor condition is achieved.
\begin{table}[h!]
\centering
\caption{Circuit parameters used in MATLAB simulation.}
\begin{tabular}{c c c}
\hline
\label{tab_sim}
\textbf{Parameters} & \textbf{Symbol} & \textbf{Value} \\ 
\hline
Grid voltage & $v_{grid}$ & 220 Vrms \\ 
Input voltage & $V_{DC}$ & 400 V \\ 
Input capacitors & $C_{in1}, C_{in2}$ & 1500 $\mu$ F \\ 
Output filter inductors & $L_{1}, L_{2}$ & 4.06 mH \\ 
Equivalent parasitic capacitor & $C_{PV}$ & 24 nF \\ 
MPPT Switching frequency & $f_{sw}$ & 20 kHz \\ 
Output power & $P_{out}$ & 2200 W \\

\hline
\end{tabular}
\end{table}

\subsection{DC-Link Analysis}
\begin{figure}[h]
    \centering
    \begin{subfigure}[b]{0.5\linewidth}
        \centering
        \includegraphics[width=\linewidth]{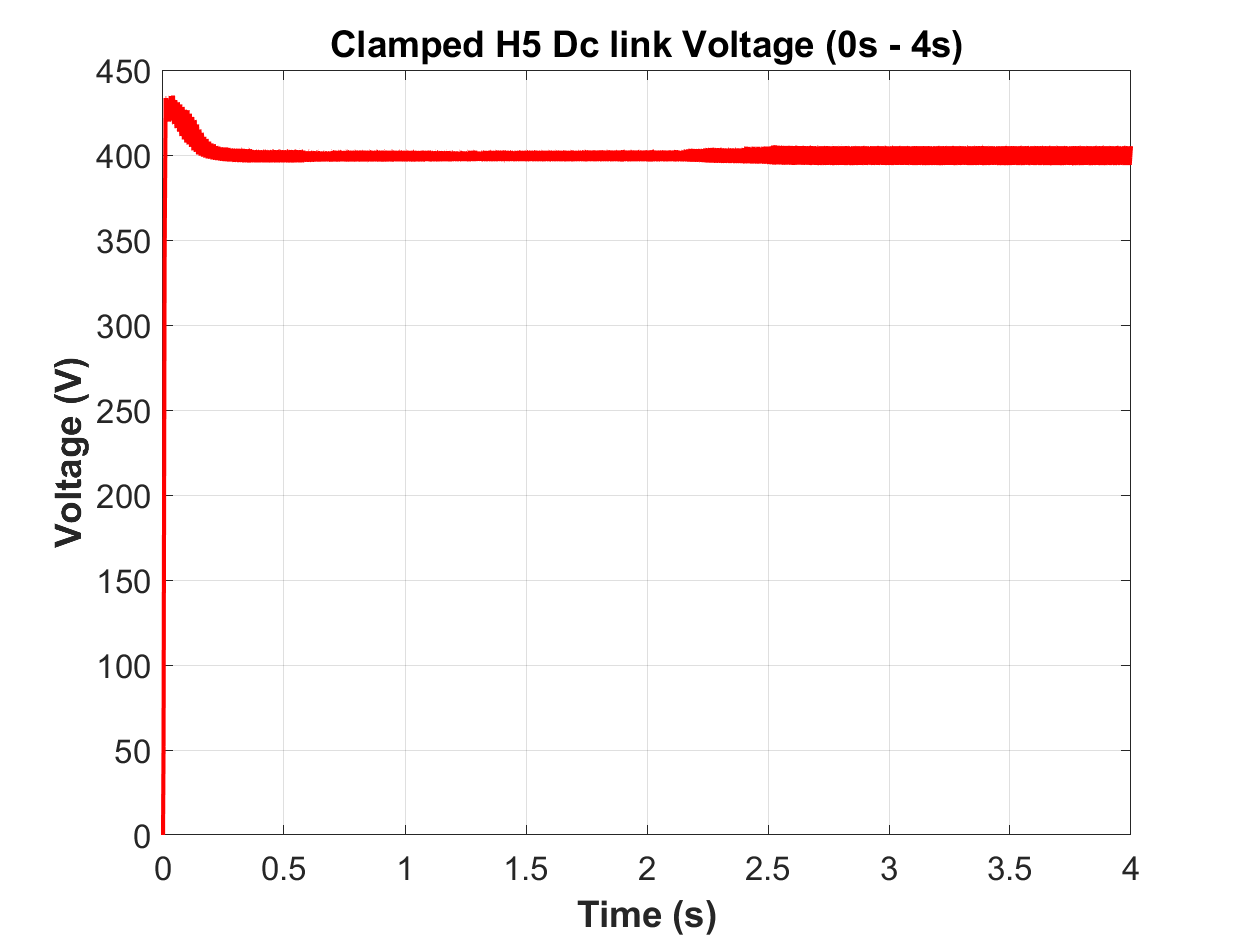}
        \caption{DC-link Voltage}
        \label{DC-link}
    \end{subfigure}%
    \begin{subfigure}[b]{0.50\linewidth}
        \centering
        \includegraphics[width=\linewidth]{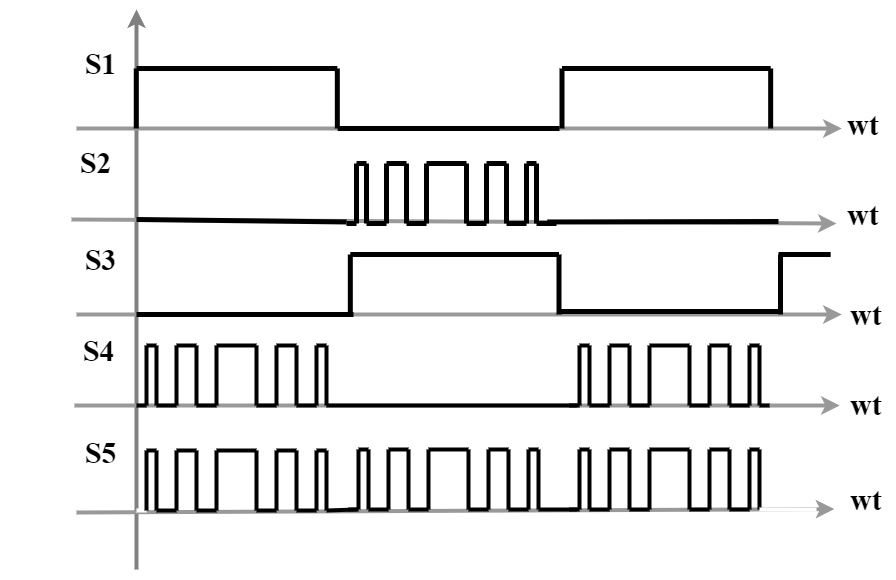}
        \caption{Switching Pattern}
        \label{fig_h5_switches}
    \end{subfigure}
    \caption{Simulation results of HCH5-D2 topology.}
\end{figure}

With the help of outer DC-link voltage control loop, the simulation model of HCHD-D2 was able to maintain the DC-link voltage at a constant value of 400V as shown in Fig.\ref{DC-link}.

Likewise, the five different switching signals for the inverter switches, S1-S5 obtained by using Inverse Park transformation of DC reference current and hysteresis band current control are shown in Fig\ref{fig_h5_switches}.

\subsection{CMV and DMV Analysis}
\begin{figure}[H]
    \centering
    \begin{subfigure}[b]{0.50\linewidth}
        \centering
        \includegraphics[width=\linewidth]{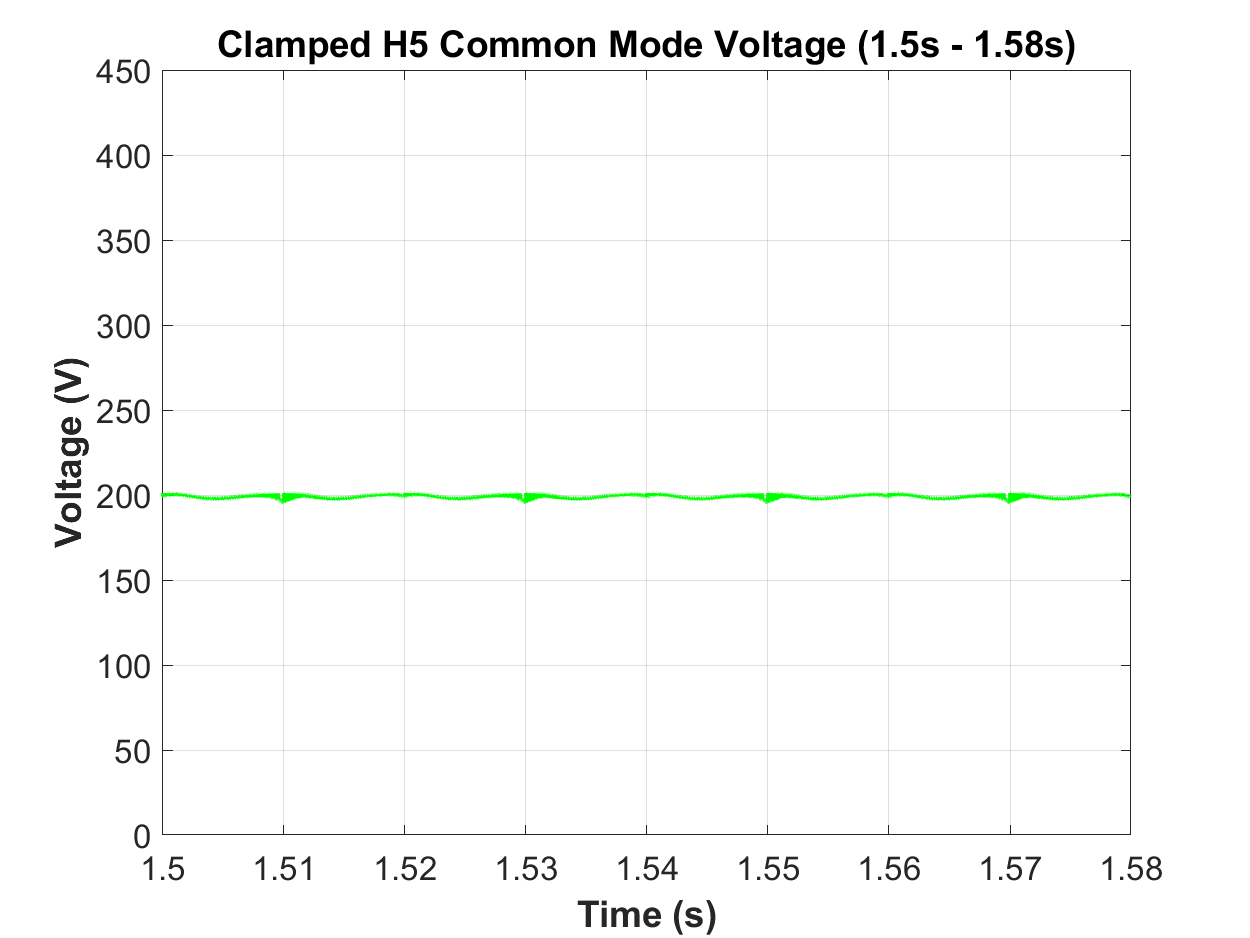}
        \caption{Common Mode Voltage}
        \label{fig_h5_cmv}
    \end{subfigure}%
    \begin{subfigure}[b]{0.50\linewidth}
        \centering
        \includegraphics[width=\linewidth]{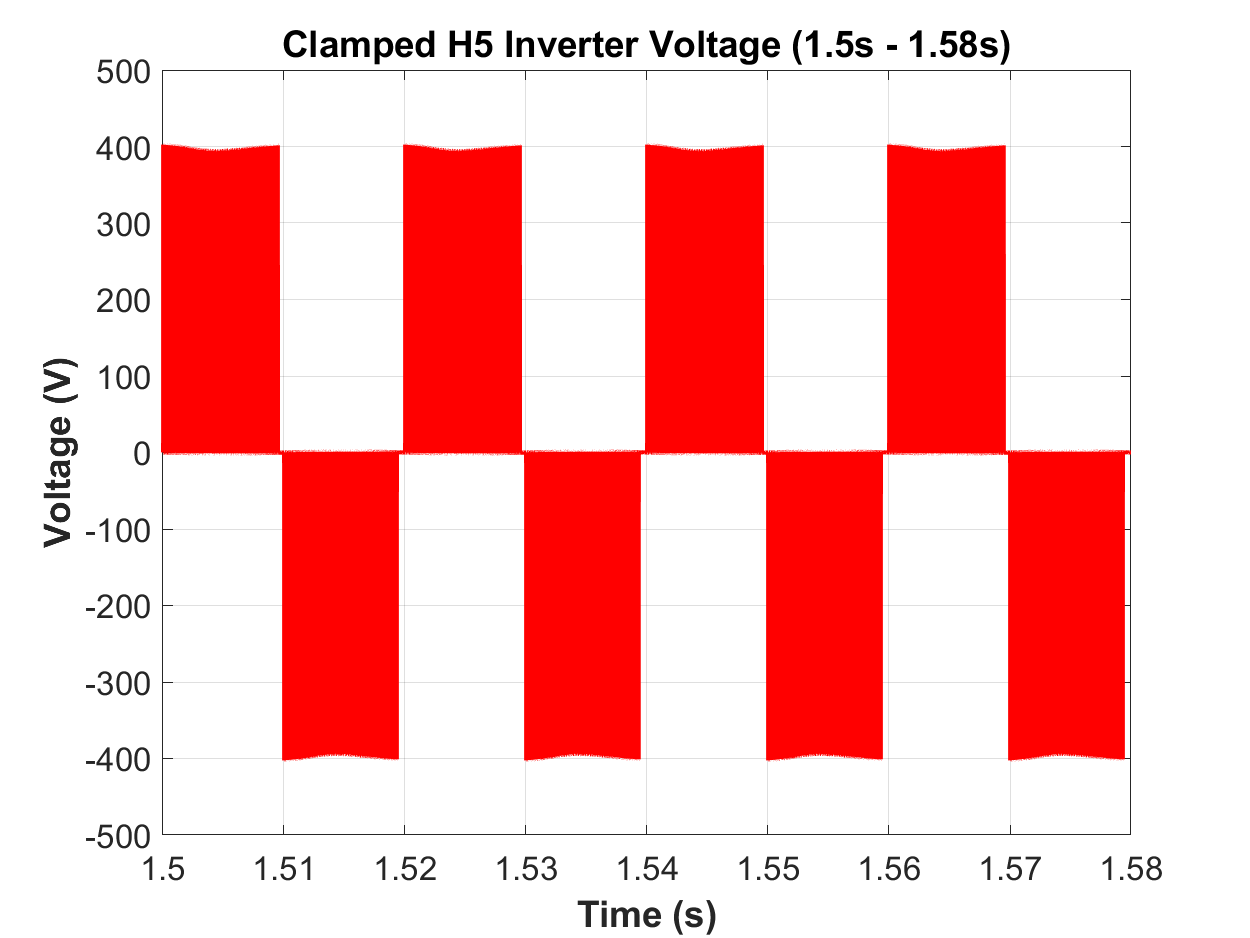}
        \caption{Differential mode voltage}
        \label{fig_h5_dmv}
    \end{subfigure}
    \caption{Simulation results of HCH5-D2 topology.}
\end{figure}

It can be seen from Fig.\ref{fig_h5_cmv}, the CMV profile is basically constant around $V_{DC}/2$ which is around 200 volts .The upper diode D1 and lower diode d2, during the freewheeling period, were able to clamp the common mode voltage at $V_{DC}/2$ and also reduces the number of spikes. Also, the DMV has unipolar nature as in Fig, \ref{fig_h5_dmv}, which ensures that the switches of the inverter are subjected to less electrical switching stress which ensures low switching losses. 

\subsection{Leakage Current Comparison}
\begin{figure}[h]
    \centering
    \begin{subfigure}[b]{0.5\linewidth}
        \centering
        \includegraphics[width=\linewidth]{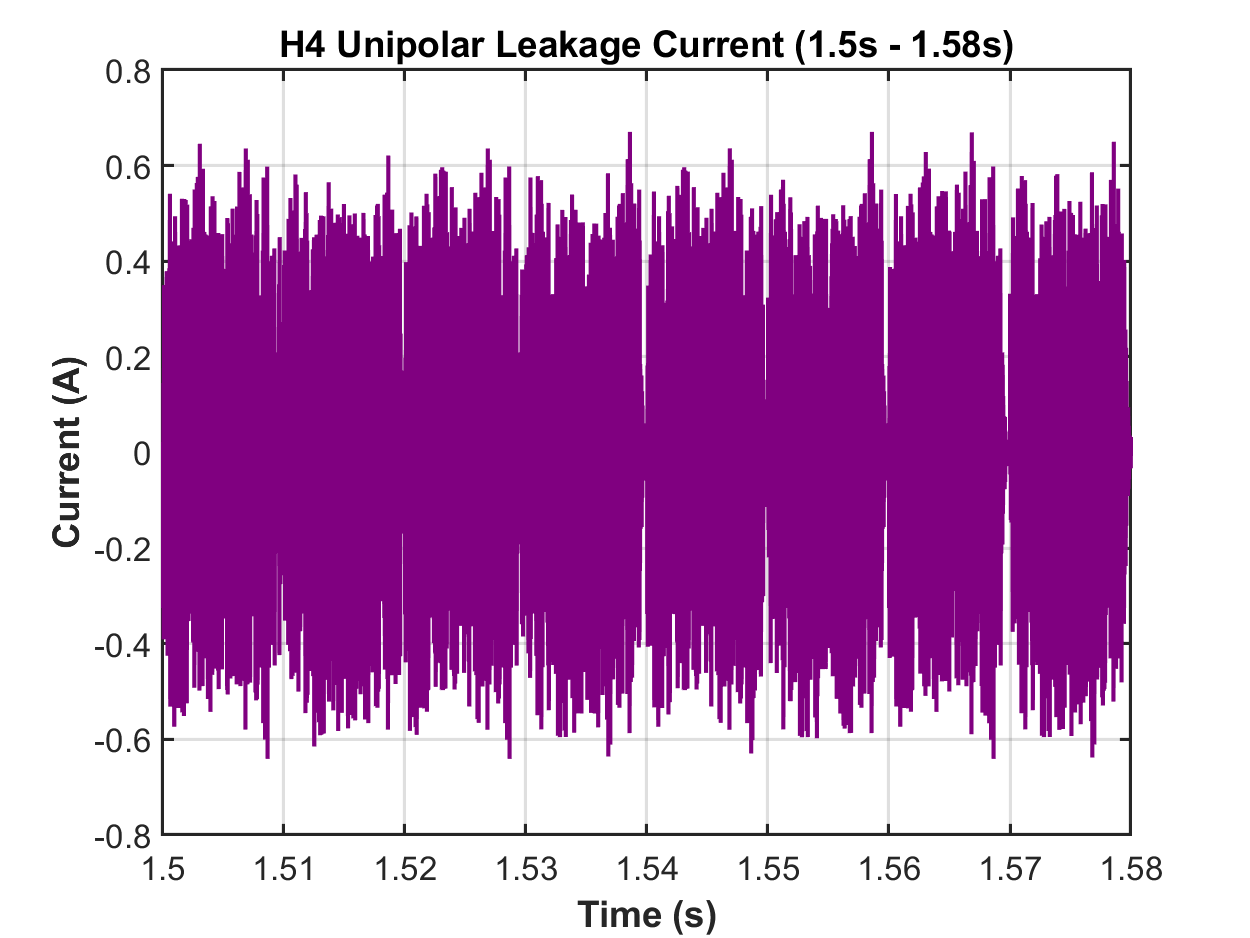}
        \caption{H4: Leakage current}
        \label{fig_leak_h4}
    \end{subfigure}%
    \begin{subfigure}[b]{0.5\linewidth}
        \centering
        \includegraphics[width=\linewidth]{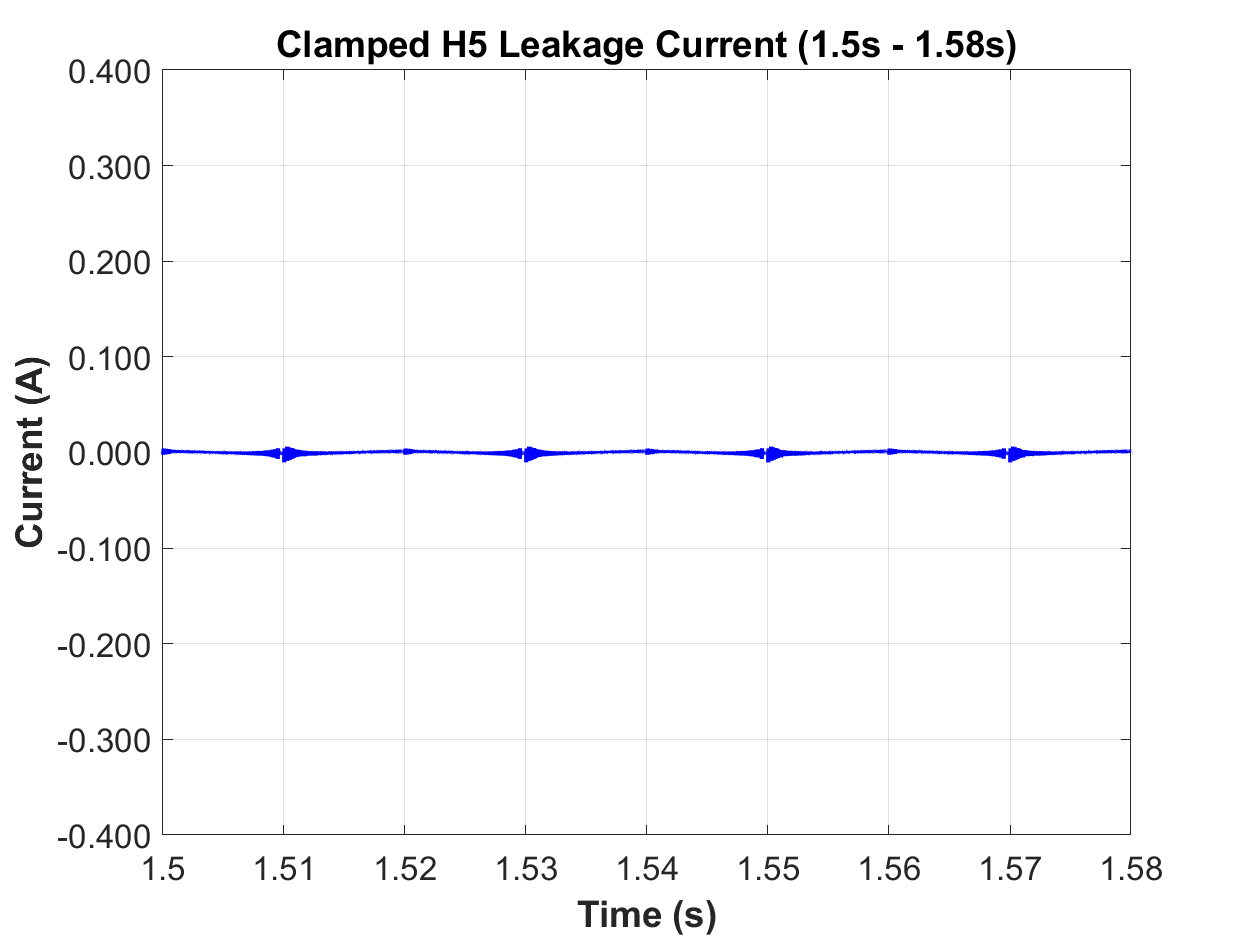}
        \caption{HCH5-D2: Leakage current}
        \label{fig_leak_h5}
    \end{subfigure}
    \caption{Comparison of simulation results: Current}
\end{figure}

The leakage current graphs of conventional H4 inverter and HCH5-D2 inverter are shown in Fig \ref{fig_leak_h4} and Fig \ref{fig_leak_h5} respectively. Comparing the maximum spikes of leakage current, which exceeded 0.6 A in the conventional H4 topology in Fig \ref{fig_leak_h4}.a, 
this magnitude of current violates the German VDE0126–1–1 standard. Moreover, there's also the possibility of triggering inaccurate signals in several protection devices including circuit breakers. However, in the HCH5-D2, the common mode current is effectively suppressed to a permissible limit by a large instant as compared to that in H4 inverter.
From Table \ref{tab:leakage_current_comparison}, it is seen that the leakage current of the HCH5D2 comes to around 1.35mA, which is significantly less than the traditional H4 bridge inverter i.e. 285mA.  

\begin{table}[h]
    \centering
    \caption{Leakage current comparison.}
    \begin{tabular}{p{3.5cm}|p{3cm}}
        \hline
        Topology & Leakage Current (RMS) \\
        \hline
        H4 with unipolar modulation & 285 mA \\
        H5 &  1.35mA \\
        \hline
    \end{tabular}
    \label{tab:leakage_current_comparison}
\end{table}
\subsection{Grid Voltage and Current}
\begin{figure}[h]
    \centering
    \begin{subfigure}[b]{0.5\linewidth}
        \centering
        \includegraphics[width=\linewidth]{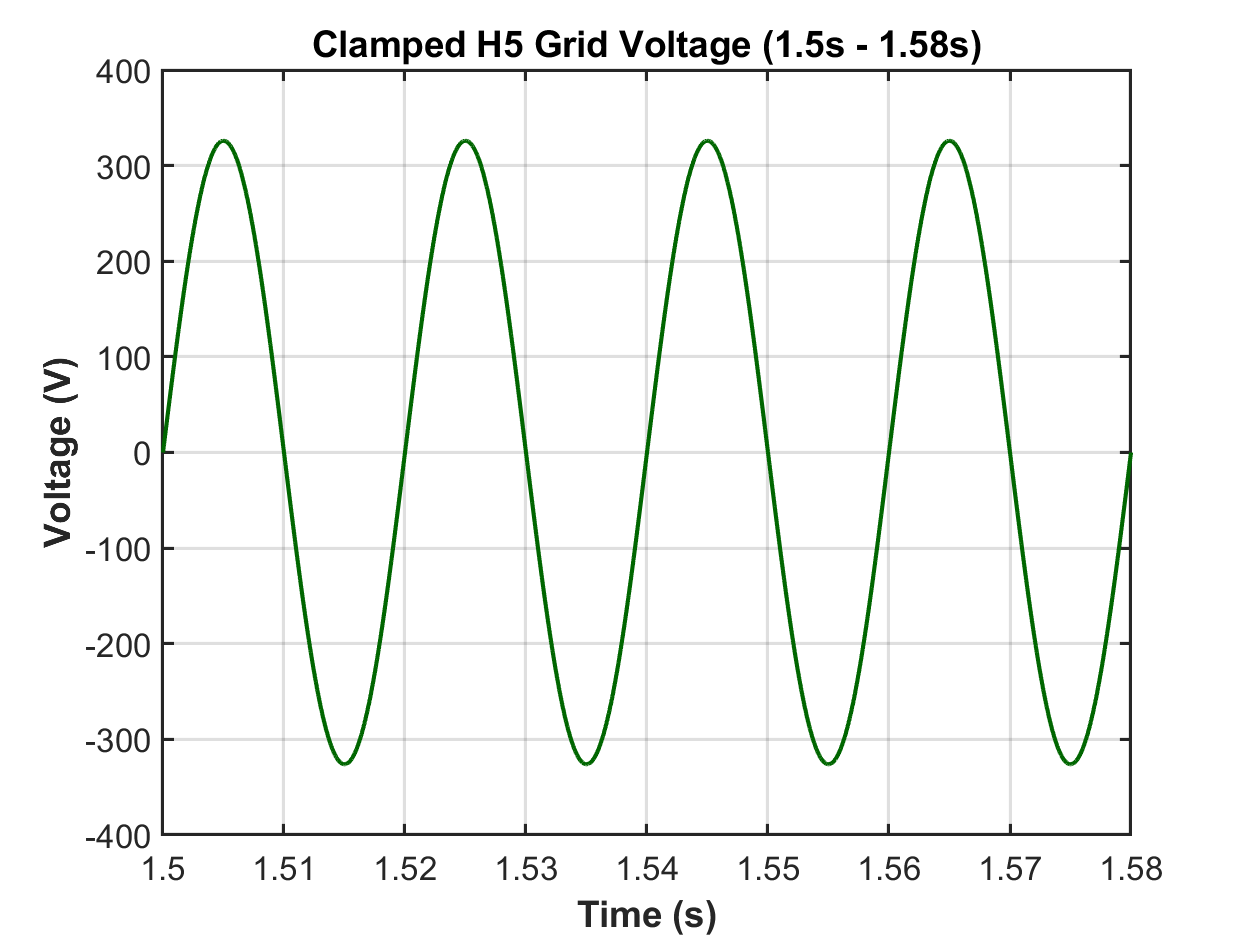}
        \caption{Grid voltage}
        \label{fig_HCH5_grid_voltage}
    \end{subfigure}%
    \begin{subfigure}[b]{0.5\linewidth}
        \centering
        \includegraphics[width=\linewidth]{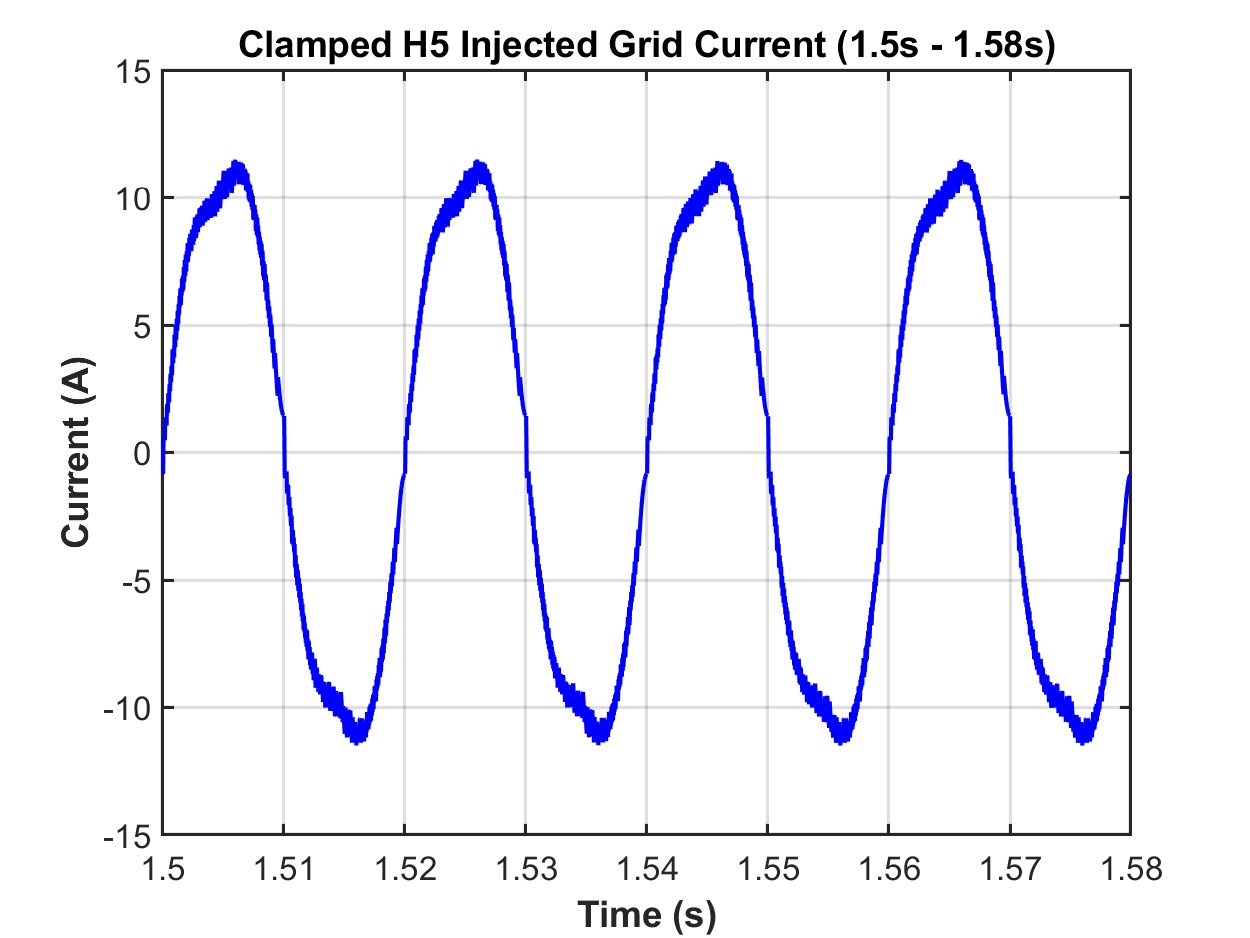}
        \caption{Grid current}
        \label{fig_HCH5_grid_current}
    \end{subfigure}
    \caption{Simulation results of HCH5-D2 topology.}
\end{figure}

The waveform of the output current showed the features of a hysteresis band like the gradual rise and decline pattern to ensure that output current follows the reference path. Similarly, the current was also synchronized with the grid voltage apart from the negligible phase lag occured due to the use of LCL filter ciruit.  These simultaneous waveforms are presented in the Fig.\ref{fig_HCH5_grid_voltage} and Fig.\ref{fig_HCH5_grid_current}. Hence, only the active power was injected into the grid.

\begin{figure}[h]
    \centering
    \includegraphics[width=0.8\linewidth]{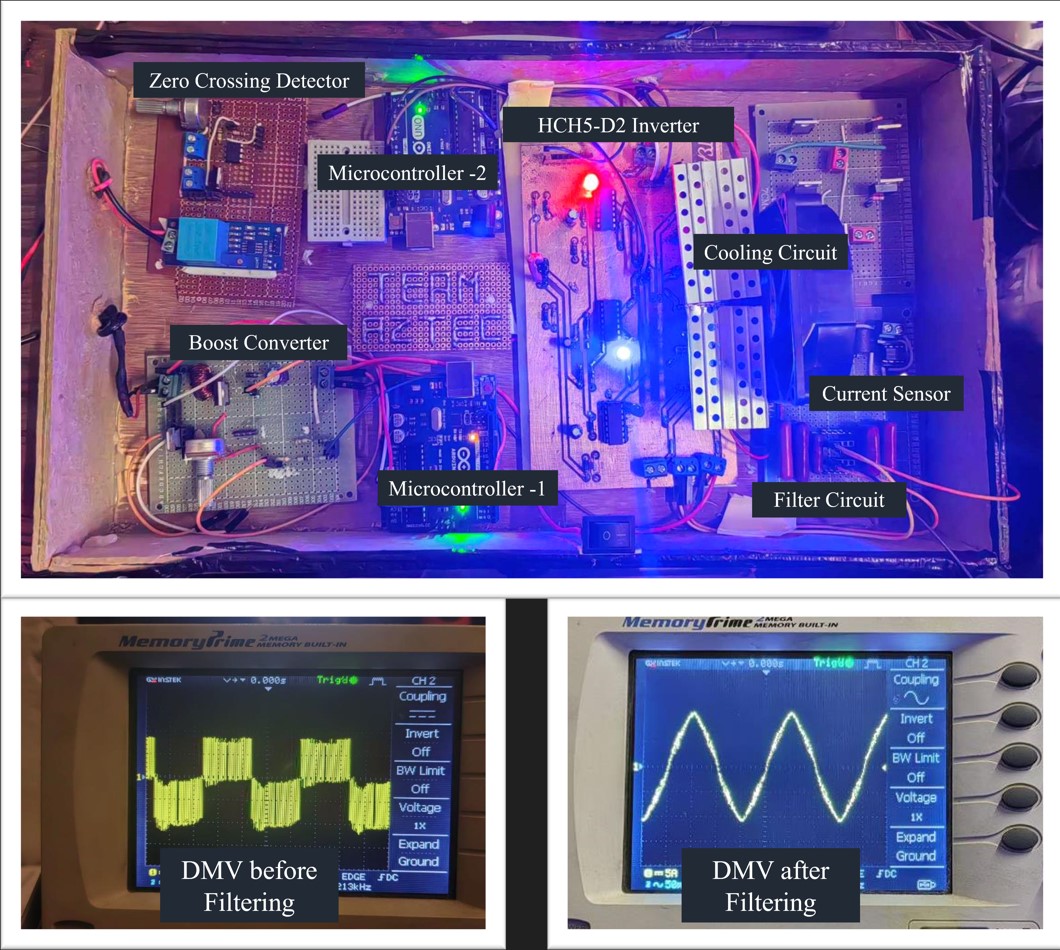}
    \caption{Experimental setup of HCH5D2 Inverter topology}
    \label{fig_expt_setup}
\end{figure}

The Fig\ref{fig_expt_setup} displays the experimental setup of the proposed HCH5D2 inverter.

    \section{Conclusion}
\label{chapter5}
A 2.2kW grid-connected single phase HCH5-D2 inverter alongside its control strategies are proposed and verified in this paper. The proposed topology was successful in maintaining constant CMV and hence significantly reducing, the leakage current in comparison to conventional H4 topology. The topology also maintains uni-polar characteristics of DMV, which ensures reduced switching stress on the inverter switches hence, minimum switching losses. The proposed topology also complies with the German VDE0126–1–1 standard. Additionally, using only five switches and hysteresis band current control this topology ensures simplicity. Hence, HCH5-D2 topology presents a scope to be a good fit for a transformer-less PV inverters, typically, in a single phase.


    \bibliographystyle{IEEEtran}

\end{document}